# Ermakov equations in quantum mechanics

Roumen Tsekov
Department of Physical Chemistry, University of Sofia, 1164 Sofia, Bulgaria

The Ermakov equation, appearing in quantum mechanics of a harmonic oscillator, is extended via dissipative and thermal terms to take into account the effect of an environment.

More than a century ago Ermakov [1] has introduced an equation, which appears in many branches of modern physics [2, 3]. In quantum mechanics the Ermakov equation comes out in the case of a harmonic oscillator [4, 5]. The evolution of the latter is usually described by the time-dependent Schrödinger equation

$$i\hbar \partial_t \psi = (-\hbar^2 \partial_x^2 / 2m + m\omega_0^2 x^2 / 2)\psi \tag{1}$$

where $m$ is the particle mass and $\omega_0$ is the oscillator own frequency. The complex wave function $\psi(x,t)$ can be presented generally in the polar form

$$\psi = \sqrt{\rho} \exp(iS/\hbar) \tag{2}$$

where $\rho$ is the probability density and $S$ is the wave function phase. Introducing Eq. (2) in Eq. (1) the latter splits in two equations [6]

$$\partial_t \rho = -\partial_x (\rho V) \tag{3}$$

$$m\partial_t V + mV\partial_x V + m\omega_0^2 x = -\partial_x Q \tag{4}$$

corresponding to the imaginary and real parts. The first equation is a continuity one and $V \equiv \partial_x S / m$ is the hydrodynamic-like velocity in the probability space. Equation (4) is a macroscopic force balance, where $Q = -\hbar^2 \partial_x^2 \sqrt{\rho} / 2m\sqrt{\rho}$ is the Bohm quantum potential. The solution of these Madelung equations $\rho = \exp(-x^2 / 2\sigma^2) / \sqrt{2\pi}\sigma$ is a Gaussian probability density, which introduced in Eq. (3) leads to an expression $V = x\dot{\sigma}/\sigma$ for the hydrodynamic-like velocity. Substituting now both expressions for $\rho$ and $V$ in Eq. (4) results in the Ermakov equation describing the evolution of the rms displacement $\sigma$

$$m\ddot{\sigma} + m\omega_0^2\sigma = \hbar^2/4m\sigma^3 \qquad (5)$$

Pinney [7] has found a general solution of the Ermakov equation. In the case of a free quantum particle ($\omega_0 = 0$) the solution of Eq. (5) is the well-known expression, describing ballistic spreading of a Gaussian wave packet, $\sigma^2 = \sigma_0^2 + (\hbar t/2m\sigma_0)^2$.

The Madelung presentation of the Schrödinger equation opens a door for introduction of dissipative forces in quantum mechanics. Usually the friction force of a particle in a classical environment depends linearly on the particle velocity. Hence, one can add a friction force $-bV$ in Eq. (4) to obtain [8, 9]

$$m\partial_t V + mV\partial_x V + bV + m\omega_0^2 x = -\partial_x Q \qquad (6)$$

where $b$ is the particle friction coefficient. Thus one arrives to dissipative Madelung hydrodynamics. Substituting the expressions for $\rho$ and $V$ in Eq. (6) yields a dissipative Ermakov equation [8-10, 3]

$$m\ddot{\sigma} + b\dot{\sigma} + m\omega_0^2\sigma = \hbar^2/4m\sigma^3 \qquad (7)$$

In the case of a strong friction one can neglect the first inertial term in Eq. (7) and the solution of the remaining equation is $\sigma^2 = (\hbar/2m\omega_0)\sqrt{1-\exp(-4m\omega_0^2 t/b)}$ [11]. It describes the relaxation of an initially excited oscillator to the ground state. In the case of a free dissipative quantum particle ($\omega_0 = 0$) this solution reduces to a known sub-diffusive law $\sigma^2 = \hbar\sqrt{t/mb}$ [11].

Equation (7) describes a dissipative quantum oscillator at zero temperature. In the case of a finite temperature the following thermal dissipative Ermakov equation is obtained [9]

$$m\ddot{\sigma} + b\dot{\sigma} + m\omega_0^2\sigma = 2\partial_\beta(1/\sigma)_b + \hbar^2/4m\sigma^3 \qquad (8)$$

where $\beta = 1/k_B T$ and the subscript $_b$ indicates that the differentiation on $\beta$ is carried out at constant friction coefficient. This equation reduces to Eq. (7) at $T \to 0$ and its equilibrium solution $\sigma_\infty^2 = (\hbar/2m\omega_0)\coth(\beta\hbar\omega_0/2)$ corresponds to the well-known expression from the quantum statistical thermodynamics. However, in the classical limit Eq. (8) differs from a well-known result. In the case of thermo-quantum diffusion Eq. (6) can be generalized to [12]

$$m\partial_t V + mV\partial_x V + bV + m\omega_0^2 x = -\partial_x (\ln\rho + \int_0^\beta Q d\beta)_b /\beta \qquad (9)$$

Substituting here the expressions for $\rho$ and $V$ yields another Ermakov-like equation which provides the correct classical limit

$$m\ddot{\sigma} + b\dot{\sigma} + m\omega_0^2 \sigma = [1/\sigma + \sigma \int_0^\beta (\hbar^2/4m\sigma^4)_b \, d\beta]/\beta \tag{10}$$

The equilibrium solution of Eq. (10) is again $\sigma_\infty^2 = (\hbar/2m\omega_0)\coth(\beta\hbar\omega_0/2)$. In the high temperature limit both equations (8) and (10) reduces to

$$m\ddot{\sigma} + b\dot{\sigma} + m\omega_0^2 \sigma = k_B T/\sigma + \hbar^2/4m\sigma^3 \tag{11}$$

The equilibrium solution of Eq. (11) $\sigma_\infty^2 = [\sqrt{1+(\beta\hbar\omega_0)^2}+1]/2\beta m\omega_0^2$ provides the correct limits at low and high temperatures. This thermal dissipative Ermakov equation is solved in the case of strong friction, where the first inertial term is neglected [13].

Finally, the Ermakov equation allows exploring other dissipative models in quantum mechanics. For instance, one can add in Eq. (5) a jerk term, heuristically corresponding to the Abraham-Lorentz force [14], to obtain

$$m\ddot{\sigma} - r\dddot{\sigma} + m\omega_0^2 \sigma = \hbar^2/4m\sigma^3 \tag{12}$$

where $r = e^2/6\pi\varepsilon_0 c^3$, $e$ is the particle charge and $c$ is the speed of light. Thus, Eq. (12) describes a charged quantum harmonic oscillator dissipating energy via electromagnetic radiation.

[1]  V.P. Ermakov 1880 *Univ. Izv. (Kiev)* **20** 1
[2]  P.G.L. Leach and K. Andriopoulos 2008 *Appl. Anal. Discrete Math.* **2** 146
[3]  F. Haas 2010 *Phys. Scr.* **81** 025004
[4]  A.B. Nassar 1985 *Phys. Rev.* **32** 1862, 1986 *Phys. Rev.* **33** 4433
[5]  D. Schuch 2008 *SIGMA* **4** 043
[6]  E. Madelung 1927 *Z. Phys.* **40** 322
[7]  E. Pinney 1950 *Proc. Am. Math. Soc.* **1** 681
[8]  A.B. Nassar 1985 *J. Phys. A: Math. Gen.* **18** L509
[9]  R. Tsekov and G.N. Vayssilov 1992 *Chem. Phys. Lett.* **195** 423
[10] A.B. Nassar 1986 *J. Math. Phys.* **27** 755
[11] R. Tsekov 2009 *Int. J. Theor. Phys.* **48** 630
[12] R. Tsekov 2009 *Int. J. Theor. Phys.* **48** 85
[13] J.A. Messer 2008 *GEB Univ. Giessen* 6328
[14] M. Abraham 1905 *Theorie der Elektrizität* (Leipzig: Teubner)